\documentclass[journal,comsoc]{IEEEtran}
\usepackage[T1]{fontenc}% optional T1 font encoding
\usepackage[utf8]{inputenc}

\usepackage{cite}
\ifCLASSINFOpdf
  \usepackage[pdftex]{graphicx}
  \graphicspath{{../pdf/}{../jpeg/}}
  \DeclareGraphicsExtensions{.pdf,.jpeg,.png}
\else
  \usepackage[dvips]{graphicx}
  \graphicspath{{../eps/}}
  \DeclareGraphicsExtensions{.eps}
\fi
\usepackage[cmintegrals]{newtxmath}
\ifCLASSOPTIONcompsoc
  \usepackage[caption=false,font=normalsize,labelfont=sf,textfont=sf]{subfig}
\else
  \usepackage[caption=false,font=footnotesize]{subfig}
\fi
\ifCLASSOPTIONcaptionsoff
  \usepackage[nomarkers]{endfloat}
 \let\MYoriglatexcaption\caption
 \renewcommand{\caption}[2][\relax]{\MYoriglatexcaption[#2]{#2}}
\fi
\usepackage{url}
\usepackage{xcolor} %TODO remove later, only for drafting..
\usepackage{siunitx}
\usepackage{graphicx}
\usepackage{glossaries}
\usepackage{booktabs} % for nicer tables
\usepackage{flushend} % flush final page properly

\makeglossaries

\newacronym{5G}{5G}{fifth generation}
\newacronym{ACC}{ACC}{adaptive cruise control}
\newacronym{CAM}{CAM}{cooperative awareness message}
\newacronym{CPCL}{CPCL}{cooperative passive coherent location}
\newacronym{CQI}{CQI}{channel quality indicator}
\newacronym{C-V2X}{C-V2X}{cellular V2X}
\newacronym{D-FFT}{D-FFT}{Doppler-FFT}
\newacronym{DoA}{DoA}{direction of arrival}
\newacronym{eNB}{eNB}{eNodeB}
\newacronym{FFT}{FFT}{fast Fourier transform}
\newacronym{IFFT}{IFFT}{inverse fast Fourier transform}
\newacronym{ITS}{ITS}{intelligent transportation systems}
\newacronym{LoS}{LoS}{line-of-sight}
\newacronym{LTE}{LTE}{long-term evolution}
\newacronym{MAC}{MAC}{medium access control}
\newacronym{MEC}{MEC}{mobile edge cloud}
\newacronym{MIMO}{MIMO}{multiple-input multiple-output}
\newacronym{MISO}{MISO}{multiple-input single-output}
\newacronym{MNO}{MNO}{mobile network operator}
\newacronym{MRP}{MRP}{market representation partner}
\newacronym{OFDM}{OFDM}{orthogonal frequency division multiplex}
\newacronym{OFDMA}{OFDMA}{orthogonal frequency division multiple access}
\newacronym{PCL}{PCL}{passive coherent location}
\newacronym{PRB}{PRB}{physical resource block}
\newacronym{QoS}{QoS}{quality-of-service}
\newacronym{RCS}{RCS}{radar cross-section}
\newacronym{RSU}{RSU}{road-side unit}
\newacronym{SC-FDMA}{SC-FDMA}{single carrier frequency division multiple access}
\newacronym{SIMO}{SIMO}{single-input multiple-output}
\newacronym{SNR}{SNR}{signal-to-noise ratio}
\newacronym{UE}{UE}{user equipment}
\newacronym{V2I}{V2I}{vehicle-to-infrastructure}
\newacronym{V2V}{V2V}{vehicle-to-vehicle}
\newacronym{V2X}{V2X}{vehicle-to-everything}
\newacronym{WSA}{WSA}{worst subcarrier avoiding}

\begin{document}

% IEEE Copyright Notice, see:
% https://www.ieee.org/publications_standards/publications/rights/rights_policies.html
\makeatletter
\def\ps@IEEEtitlepagestyle{
    \def\@oddfoot{\mycopyrightnotice}
    \def\@evenfoot{}
}
\def\mycopyrightnotice{
    {\scriptsize
    \begin{minipage}{\textwidth}
    \centering
    © 2019 IEEE. Personal use of this material is permitted.
    Permission from IEEE must be obtained for all other uses, in any current or future media, including reprinting/republishing this material for advertising or promotional purposes, creating new collective works, for resale or redistribution to servers or lists, or reuse of any copyrighted component of this work in other works.
    \end{minipage}
    }
}

\title{Cooperative Passive Coherent Location: \protect\\ A Promising 5G Service to Support Road Safety}


\author{
    \IEEEauthorblockN{
        Reiner~S.~Thom\"a,~\IEEEmembership{Fellow,~IEEE}, %\IEEEauthorrefmark{1}\IEEEauthorrefmark{2},
        Carsten~Andrich, %\IEEEauthorrefmark{2},
        Giovanni~Del~Galdo,~\IEEEmembership{Member,~IEEE}, %\IEEEauthorrefmark{1}\IEEEauthorrefmark{2},
        Michael~D\"obereiner, %\IEEEauthorrefmark{1},
        Matthias~A.~Hein,~\IEEEmembership{Senior Member,~IEEE}, %\IEEEauthorrefmark{1},
        Martin K\"aske, %\IEEEauthorrefmark{1},
        G\"unter~Sch\"afer,~\IEEEmembership{Member,~IEEE}, %\IEEEauthorrefmark{1},%TODO ue/ae?
        Steffen~Schieler, %\IEEEauthorrefmark{1},
        Christian~Schneider, %\IEEEauthorrefmark{1},
        Andreas~Schwind, %\IEEEauthorrefmark{1}
        and~Philip~Wendland %\IEEEauthorrefmark{1}
    }\\
%    \vspace{0.5em}
    \thanks{All authors are with the Technische Universit\"at Ilmenau (Ilmenau University of Technology). Carsten Andrich, Michael D\"obereiner, and Giovanni Del Galdo are with the Fraunhofer IIS.

Published in IEEE Communications Magazine, vol.~57, no.~9, pp.~86--92, September 2019.
DOI: 10.1109/MCOM.001.1800242

This work was supported by the Freistaat Thüringen and the European Social Fund.
    }% <-this % stops a space
%\thanks{Manuscript received April XX, 2018; revised XXXXX XX, XXXX.}
%    \IEEEauthorblockA{\IEEEauthorrefmark{1}Ilmenau University of Technology}\\
%    \IEEEauthorblockA{\IEEEauthorrefmark{2}Fraunhofer IIS}
}

\maketitle

\begin{abstract}
5G promises many new vertical service areas beyond simple communication and data transfer.
We propose CPCL (cooperative passive coherent location), a distributed MIMO radar service, which can be offered by mobile radio network operators as a service for public user groups.
CPCL comes as an inherent part of the radio network and takes advantage of the most important key features proposed for 5G.
It extends the well-known idea of passive radar (also known as passive coherent location, PCL) by introducing cooperative principles.
These range from cooperative, synchronous radio signaling, and MAC up to radar data fusion on sensor and scenario levels.
By using software-defined radio and network paradigms, as well as real-time mobile edge computing facilities intended for 5G, CPCL promises to become a ubiquitous radar service which may be adaptive, reconfigurable, and perhaps cognitive.
As CPCL makes double use of radio resources (both in terms of frequency bands and hardware), it can be considered a green technology.
Although we introduce the CPCL idea from the viewpoint of vehicle-to-vehicle/infrastructure (V2X) communication, it can definitely also be applied to many other applications in industry, transport, logistics, and for safety and security applications.
\end{abstract}

\begin{IEEEkeywords}
5G verticals, vehicle-to-x (V2X), cooperative driving, intelligent transport systems (ITS), joint communication and radar, passive coherent location (PCL), passive OFDM radar, distributed MIMO radar network, radar resource management, high-resolution radar parameter estimation
\end{IEEEkeywords}

% \fi
\IEEEpeerreviewmaketitle

\section*{Introduction}

The \gls{5G} mobile communication networks will be driven by several key enabling technologies \cite{AKYILDIZ201617}.
Among these are software-defined adaptivity and resource allocation on the radio and network layers, massive \gls{MIMO}, new frequency bands and waveforms, device-to-device connectivity, and so on.
This, together with low latency communication and \gls{MEC} computing, will open new horizons in service delivery.
We will observe a transformation of radio networks from pure wireless connectivity to a network for services, which will foster new fields, use cases, and business models for vertical industry applications.

Many of these, including automotive, industrial automation, and security tasks, will need location services.
Whereas positioning of mobile devices and objects provided with wireless tags is already widely discussed, there is an increasing demand for positioning of objects that are not equipped with any specific technical means to determine and report their location.
Obviously, this task requires radar location principles, which rely on proper radio illumination of the objects of interest and sensing of the backscattered signals.
Here, we propose the new principle of \gls{CPCL}, which is to be an integrated radar service of future mobile radio networks.
 Essentially, \gls{CPCL} extends the well-known idea of passive radar, also known as \gls{PCL}.
Whereas \gls{PCL} does not consider any cooperation between radar illuminators and sensors, we assume for \gls{CPCL} that all radar nodes belong to the same network.
This way, \gls{CPCL} will turn the mobile radio network into a distributed \gls{MIMO} radar network, which opens a wide scope of cooperation between sensor nodes reaching from cooperative bi-/multi-static target scene illumination up to radar data networking and fusion.

Although it seems to be applicable to different mission-critical vertical services in automotive, logistics and public safety, we introduce the \gls{CPCL} idea from the viewpoint of cooperative driving.
Therefore, this article starts with a short and concise overview of \gls{V2V}/infrastructure communications (V2X) with an emphasis on \gls{LTE} \gls{V2X} and the \gls{5G} perspective.
We give a survey of the current situation of automotive radar as one location sensor principle for automated and cooperative driving and review conventional \gls{PCL}.
Based on this, we elaborate on the basic idea of \gls{CPCL}, highlight the challenges and the potential of \gls{CPCL} as an inherent radar service in future \gls{5G} networks, and summarize the most important related research questions.
We also give a first measured example to demonstrate its feasibility.

\section*{Current Situation in \gls{V2X} Communications and Radar Sensing}
\gls{CPCL} builds upon various technologies and developments in wireless mobile and vehicular communication networks, as well as traditional radar sensing approaches.

\subsection*{5G Perspective for \gls{V2X} Communications}
With the LTE-V2X standard, 3GPP recently has made the next step towards \gls{5G} \gls{V2X} communications, accelerated by the global cross-industry \gls{5G} Automotive Association as a cooperating \gls{MRP}.
This alliance proposes the coexistence of \gls{C-V2X} and ITS-G5 by spectrum sharing \cite{5gaa-whitepaper}.
Given the virtual ubiquity of cellular infrastructure, \gls{C-V2X} will enjoy all advantages of a commercial cellular network managed by \glspl{MNO}.
The \gls{V2X} roadside access could be handled with the same field equipment that is rolled out for cellular services.
It provides a scalable and extendable technology platform and paves the road to \gls{5G}. This allows for quality-of-service control and offers seamless network access to all resources, services, and contents offered by the \glspl{MNO}.
Moreover, \glspl{MNO} can define and offer specific services for road users and schedule radio and network resources according to their needs.

\subsection*{Road Traffic Situation Awareness and Cooperative Radar Sensing}
The visionary aims of ITS-G5 are automated and connected driving, road safety, and traffic efficiency.
Communication between cars and the dedicated infrastructure in terms of messaging is one enabler for cooperative driving.
The \glspl{CAM} enable gaining road traffic situation awareness in real time.
However, it is restricted to appropriately equipped entities and relies on self-location of cars based on satellite and inertial navigation and map matching.
While the automobile industry still uses the term “ego-car” to emphasize the autonomy and self-reliance of the car driver, it becomes obvious that more advanced cooperation can significantly enhance road-traffic situation awareness.
Efficient control and coordination of a certain traffic situation on intersections would require centralized data processing that collects information from all sensors carrying entities and fuses it with additional information available in the \glspl{RSU} from auxiliary sensors or from databases (like maps).

Radar sensors are very well established for \gls{ACC} and collision avoidance.
They range over long distances and under bad weather conditions, do not need visible light illumination, allow direct relative speed measurements, and provide overview coverage.
However, the current penetration rate is still low and radar application mostly restricted to high-end cars and trucks.
Future cars will have multiple radar systems on board to extend the field of view and the duty cycle will increase to cope with highly dynamic scenarios.
It is more than reasonable to predict an exponential growth in radar sensor density.
However, massive radar sensing will cause a lot of interference and interoperability problems \cite{7733011}.
The core problem of coexistence is that existing automotive radar does not include any advanced medium access control scheme.
There are also only limited possibilities to provide new frequency bands for hosting more radar users since even in the millimeter-wave frequency region there is already an increasingly strong competition with communication systems.

\subsection*{Overview of \gls{PCL}}
Radar has a long history in military and civilian air, space, and maritime surveillance.
Although there are several parallels, radar and radio communications have developed separately in history and radio resources (frequency bands) are typically used in an exclusive and sometimes competitive way -- with the remarkable exception of \gls{PCL}.
Passive radar does not use a dedicated transmitter for target illumination.
Instead, \gls{PCL} uses so-called transmitters-of-opportunity, for example, terrestrial broadcast transmitters or cellular communication systems.
Obviously, range, coverage, and resolution scale with transmit power and bandwidth, which makes the applicable primary radio source dependent on the required target location performance.
A topical overview on passive radar for civilian and military application is given in \cite{7917418}.

The basic \gls{PCL} setup consists of a transmitter-of-opportunity illuminating the target and a dedicated remote \gls{PCL} sensor receiving both the \gls{LoS} signal and the signal that is backscattered from the target.
The former is taken as a reference and being correlated with the backscattered wave.
The excess delay derived from the correlation maximum defines an ellipse, describing the possible positions of the target relative to the transmitter and the sensor positions.
%Therefore, the reference link between the observing and the illuminating node should be \gls{LoS} to ensure the received reference signal suits as a geometrical reference, should have a high \gls{SNR}, and should not be distorted by multipath propagation.
Another pair of nodes (e.\,g., a second transmitter) would provide another ellipse, with their intersections indicating the potential positions of the target.
This reveals a basic difference between \gls{PCL} and standard automotive radar.
The latter is ”monostatic” meaning that the target is illuminated and observed by the same antenna, or by antennas that are almost at the same position (quasi-monostatic).
The illumination and observation geometry of \gls{PCL} is called bi-static or multi-static (in case of multiple illuminators and/or observers), also known as distributed \gls{MIMO} radar \cite{4408448}.
As \gls{PCL} relies on the ubiquitous available broadcast or cellular radio transmitters, it neither needs dedicated transmitters, nor additional frequency resources.

\section*{\gls{CPCL}: Basic Concept}
Given the described state of the art, let us ask: Can passive radar become an inherent service of a mobile radio network?
As with \gls{PCL}, \gls{CPCL} makes double use of the communication signals.
Contrary, however, in \gls{CPCL} the sensors are not independent from the mobile communication network.
The radar nodes are booked in as \gls{UE} devices.
This offers various opportunities for cooperation between radar-\glspl{UE} and the network.
In \gls{CPCL}, any radio node can act as an illuminator or observer.
In traffic scenarios, this may include vehicles and fixed illuminators, like \glspl{RSU} or base stations (\gls{eNB}).
In such a joint communication and radar network, cooperation has many facets.
On the signal level, \gls{CPCL} can profit from synchronization of all radio nodes involved, maintaining mutual orthogonality.
Additionally, medium access control mechanisms minimize congestion, interference, and collisions.
Upcoming \gls{V2X} communications will inherently submit cooperative vehicle status information such as precise position and speed required for radar location and Doppler reference.
Radio nodes can further cooperate by adjusting their radio resource parameters according to target location needs.
Locally estimated target parameters can be exchanged by the same radio network and fused on a higher level.
Generally speaking, \gls{CPCL} can make use of all the network resources to be deployed in \gls{5G}, creating an unprecedented, powerful radar network.

Another facet of \gls{CPCL} is its inherent resource efficiency.
It makes dual-use of the allocated scarce frequency resources.
Also, the radio system resources are dual-used, reducing costs.
Moreover, with the mobile radio network, an ubiquitous radar service will be available whose coverage and performance automatically improve with future updates to network resources.

\begin{figure}[!t]
    \centering
    \includegraphics[width=.9\linewidth]{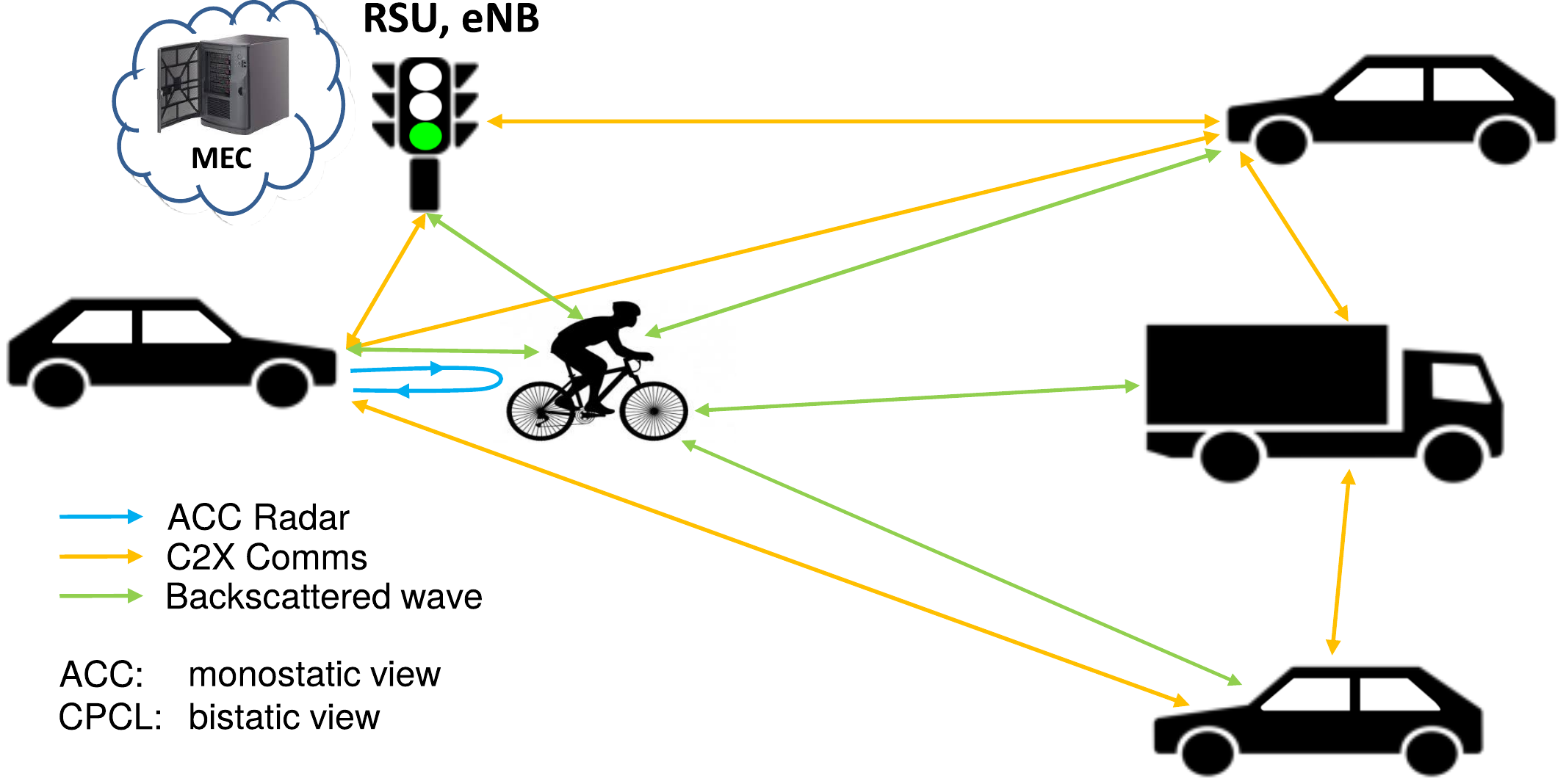}
    \caption{CPCL – Traffic Scenario.
\gls{V2X} communication signals of cars and roadside units (yellow arrows) illuminate the road user including those not equipped with \gls{V2X}.
All information available from the backscattered signals (green arrows), other sensors (such as \gls{ACC} radar) and a priori information, for example, from maps, is processed together in the \gls{MEC}, allowing for low latency computational services.}
    \label{fig_cpcl}
\end{figure}

\subsection*{5G Key Enabling Technologies Relevant for \gls{CPCL}}
The features announced for \gls{C-V2X} (see 3GPP release 14) already widely support the \gls{CPCL} idea.
This holds even more for future \gls{5G} networks.
The scalable radio access techniques based on \gls{OFDMA} and the upcoming generalized and filter bank based versions \cite{AKYILDIZ201617} are nicely suited for radar signal processing.
\gls{V2X} would support different \gls{MIMO} radar setups where any radio node like the ones depicted in Fig. \ref{fig_cpcl} can act as a radar node.
This includes \gls{SIMO} (from downlink communication), \gls{MISO} (from uplink communication), and \gls{MIMO} (from device-to-device and \gls{V2V} communication).
Channel bonding and carrier aggregation can deliver increased bandwidth and frequency diversity for enhanced range estimation.
Even more bandwidth for high range-resolution will be available at millimeter-wave frequencies.
If full-duplex transceivers appear, \gls{CPCL} gains an extra monostatic property.
Massive array beamforming will allow high-resolution spatial (directive) filtering and estimation.
Finally, low latency communication and powerful computing resources in the \gls{MEC} support real-time interaction between cars and infrastructure, data fusion, and controlling radar PHY-parameters in road traffic environments \cite{7736181}.

\section*{CPCL Challenges in Signal Processing, Data Fusion and Implementation}
Although \gls{OFDM} has been used as a wideband excitation signal for channel sounding for many years \cite{cs-eurasip}, \cite{843078}, it was only recently considered a favorable radar waveform.
Moreover, \gls{OFDM} is the native illumination waveform in case of \gls{PCL} together with DVB-T, DAB, WLAN or \gls{LTE} \cite{5393298}.
From frequency domain system identification, it is well known that periodic multi-frequency signals guarantee a leakage-free computation of the signal spectra through \gls{FFT}, which stands for a low estimation variance of the frequency response function.
This assumes that a cyclic prefix is applied and carrier orthogonality is maintained at the receiver.
So, for \gls{OFDM} the basic assumptions of optimum signal processing in communications and radar coincide.

\begin{figure}[!t]
    \begin{center}
    \includegraphics[width=.9\linewidth]{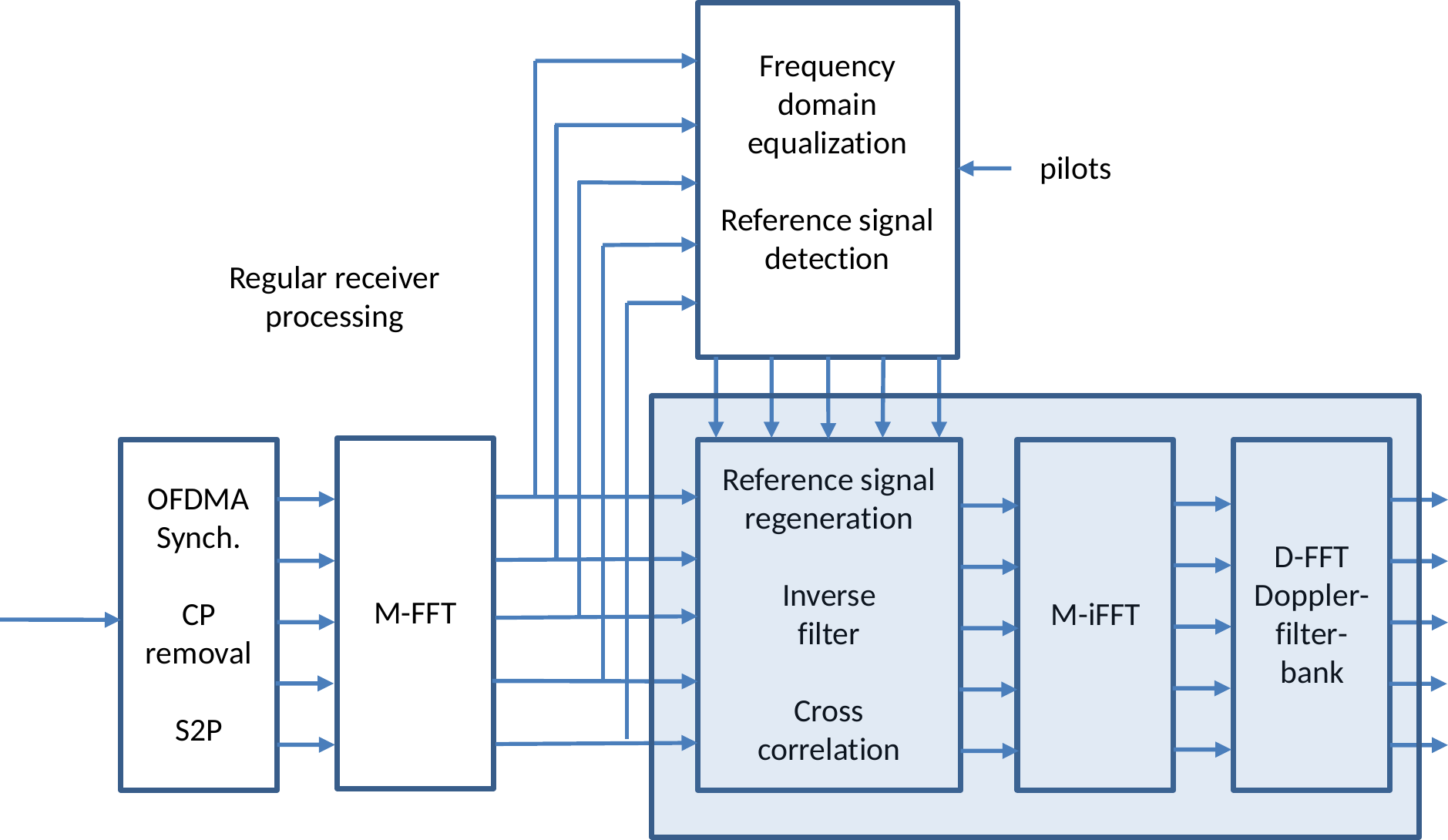}
    \end{center}
    \caption{\gls{OFDM} signal processing scheme for estimation in the joint delay-Doppler domain. The non-marked blocks correspond to regular receiver processing, whereas the marked blocks describe the radar specific processing which consists of normalized cross-correlation, \gls{IFFT} for channel impulse response calculation and another \gls{FFT} filter bank to transform to the Doppler domain (S2P: serial to parallel conversion,  M: number of carriers, D: number of \gls{OFDM} symbols used for Doppler filtering). This results in the M$\times$D two-dimensional delay-Doppler spreading function.}
    \label{fig_ofdm_proc}
\end{figure}

As a communication signal is modulated by the information data stream, we do not a priori know the transmit signal, which is needed as a correlation reference for radar signal processing.
Fortunately, in a cooperative communication environment, all the advanced measurements for robust signal reception for modern mobile radio, can be applied for transmit signal recovery.
\gls{CPCL} does not need an auxiliary reference receiver channel.
Figure \ref{fig_ofdm_proc} shows the basic receiver signal flow.
\gls{OFDM}-based \gls{CPCL} includes the standard signal processing chain of synchronization, cyclic prefix removal, \gls{FFT}, channel estimation, and cyclic frequency domain equalization.
Once the transmitted symbol is recovered, the channel frequency response function is calculated symbol-wise by inverse filtering.

Subsequent \gls{IFFT} results in the channel impulse response which indicates the multipath time delay.
In radar terminology, the time delay is called "fast time".
This type of channel estimation is different from the one required for data transmission.
Radar needs higher dynamic range (as we are looking for small details in the impulse response) and maximum rate consecutive channel impulse response processing for Doppler shift estimation and Doppler filtering.
This filtering is implemented as another \gls{FFT} filter bank (\gls{D-FFT}) along the so-called "slow time" axis, which describes the temporal evolution of the impulse response.
Hereby, we assume that the channel response factorizes with respect to time-delay and Doppler frequency, which requires the \gls{OFDM} symbols to be shorter than the channel coherence time.
This is usually the case in mobile radio.

Target detection will be carried out in the magnitude-squared delay-Doppler spreading function, which is known as scattering function.
The respective maximum integration time, which corresponds to the slow time \gls{D-FFT}-window, is limited by the moving speed of the target and the respective change in delay relative to the width of any delay bin on the fast time axis.
Maximizing the integration time allows \gls{SNR} gain and, hence, better detection of weaker target returns.
This type of Doppler shift processing is the key for separating the signals scattered back from the moving targets from those of the static environment.

\begin{figure}[!t]
    \centering
    \includegraphics[width=.9\columnwidth]{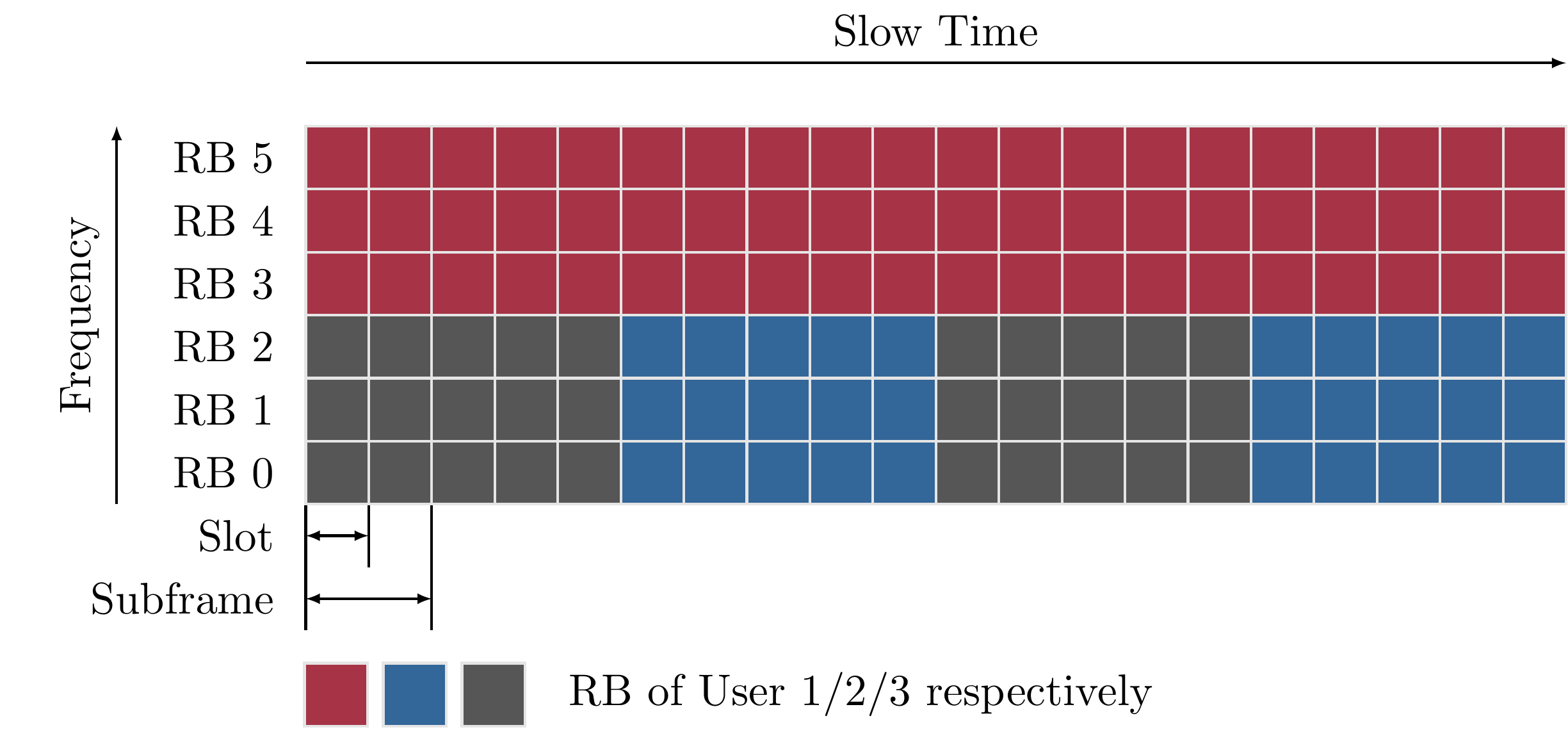}
    \caption{Example \gls{LTE} resource allocation (frequency-time resource grid) of three users (marked by different colors). The physical resource blocks are composed of 7 \gls{OFDM} symbols with 180 kHz bandwidth and contain reference symbols.}
    \label{fig_rb}
\end{figure}

A specific problem arises from the multi-user resource allocation in \gls{LTE} in frequency and time as illustrated by Fig. \ref{fig_rb}.
If the resource grid would be occupied uniformly by \glspl{PRB} belonging to a single user only, the magnitude-squared ambiguity function would be sinc-squared in the range and Doppler domains.
However, in the multi-user case, the \glspl{PRB} for any user are distributed more or less sparsely and multiple users are interleaved in frequency and time.
There can even be blanks.
In case of a downlink-radar (eNB to \gls{UE}), any radar-\gls{UE} could perhaps process the full \gls{OFDMA} symbol.
For an uplink-radar (UE to \gls{eNB}) there is no such chance.
The parts of the radio frame, which belong to different \glspl{UE} in the up-link, have to be considered as a separate measurement.
Hence, the delay-\gls{IFFT} has to process the \glspl{PRB} belonging to different radar-\glspl{UE} separately.

The resulting sparse occupation in the frequency-time plane would degrade the shape of the resulting ambiguity function.
Hence, more sophisticated range-Doppler parameter estimation procedures are required.
One option would be sparse reconstruction based on compressive sensing schemes.
Another one is model-based parameter estimation, for example, as described in \cite{cs-eurasip, 6060881, 7833233}.
The latter needs a physically motivated parametric data model to represent both the multipath propagation as well as the instrument function of the device signal processing chain, which can be determined by calibration.
This data model would effectively interpolate the missing samples in the sparse frequency-time resource grid and extrapolate it allowing for high resolution in the delay/Doppler plane beyond Rayleigh resolution.

However, there is a tremendous amount of open research questions.
Here we can only provide a short overview.

\subsection*{Radar Signal Processing}
\gls{CPCL} represents a distributed \gls{MIMO} radar network in which the illuminator, the sensor, as well as the target can be moving.
This means that the clutter originating from the static environment may now have a Doppler shift.
The separation of target signal returns from clutter may be enhanced by estimating target tracks.
Target tracking can be supported if dynamic target parameters such as speed vectors beyond mere instantaneous location are estimated.
To keep up with the fast changing scenario, the Doppler-FFT needs to be replaced by a recursive estimator to reduce latency time.

Moreover, spatial, frequency, and temporal (along slow time) diversity can be exploited to further enhance detection probability.
Spatial and frequency resources can be locally concatenating or widely distributed to maximize diversity gain and/or resolution.
The distributed respectively multi-static radar geometry gains spatial diversity because of the inherent aspect-angle variability of the bi-static \gls{RCS}. On the other hand, multiple co-located antennas (respectively antenna arrays) at eNB and even UEs would allow spatial filtering and directional estimation.

The complementary counterpart in terms of frequencies is variable bandwidth, bonding of neighboring frequency channels, or aggregation of widely fragmented bands.
The concatenated bands may be mutually coherent or non-coherent. High range resolution can be expected even at lower frequencies (hence lower bandwidth) if diverse sub-bands are available.

Therefore, \gls{CPCL} radar requires advanced distributed detection and estimation schemes.

\subsection*{Network, Signaling, Synchronization, and Hardware Issues}
There are many research questions related to radio design, such as maximizing the receiver dynamic range as weak radar echoes have to be identified in the presence of strong \gls{LoS} and clutter signals.
Massive array beamforming will support \gls{LoS} reference signal extraction by multipath filtering, relax dynamic range problems, and allow \gls{DoA} estimation for target location.
The millimeter-wave bands envisaged for \gls{5G} will offer bandwidths comparable to those of current automotive radars.
In terms of the radio network, \gls{CPCL} could use both \glspl{eNB} or \glspl{UE} as illuminators, resulting in \gls{SIMO} or \gls{MISO} radar networks.
Direct cooperation of multiple \glspl{eNB} or \glspl{UE} would allow building \gls{MIMO} radar networks.
Optimum design rules and achievable performance figures are unknown at present.
This holds true if we compare upcoming \gls{5G} and ITS-G5 for the case of \gls{V2V}.
Fully synchronous, \gls{eNB} controlled, and inherently parallel operation of multiple moving radar \glspl{UE} in \gls{5G} will be a big advantage.

\subsection*{Communication vs. Radar Resource Scheduling}
\gls{CPCL} will develop its highest potential if the radio resources would be allocated and managed in a suitable way to adapt and optimize the joint radar and communication performance.
This would include choosing the proper \gls{PRB}-distribution in time and frequency, allocation of multiple radio bands, predistortion, and allocation of spatial resources.
For instance, well-known capacity maximizing \gls{OFDM} subcarrier power allocation schemes like water filling and \gls{WSA} algorithms have already found their equivalence in multiple and extended target \gls{MIMO} radar \cite{4200705}.
Specific procedures will be applicable if spatial precoding is involved.
Without centrally controlled resource scheduling, for example, in 802.11p or \gls{LTE}-V (in case of missing cellular coverage), distributed \gls{MAC} mechanisms would need to coordinate radar and communication resources accordingly.

\subsection*{Data Fusion and Adaptive Operation}
\gls{CPCL} inherently is a multi-sensor technology.
This means that a wide variety of measurements is available, which have different uncertainty characteristics.
The key estimation procedures will rely on Bayesian data fusion, multiple hypothesis estimation, and tracking \cite{koch2016tracking}.
This requires different levels of data fusion ranging from fusion of local platform data to distributed fusion, and dynamic scene analysis at critical traffic hotspots.
Real-time map services will submit precise location information of static objects usable as reference landmarks for \gls{CPCL} calibration.
The use of the real-time computing facilities of the \gls{MEC} for \gls{CPCL} distributed data fusion, sensor resource allocation, and sensor mission control will be a challenging field of research.
We believe that reusing of the communication payload signals as radar illumination signals, instead of defining specific “radar pilots”, gives us the required signal design flexibility for wide scale radar system performance adaptation.
The \gls{MEC} also bridges the gap between the “local awareness bubble” to the higher geographical layers of traffic control.

\section*{Initial Measured Example}
\begin{figure}[!t]
    \centering
    \subfloat[]{
        \centering
        \includegraphics[width=.95\columnwidth]{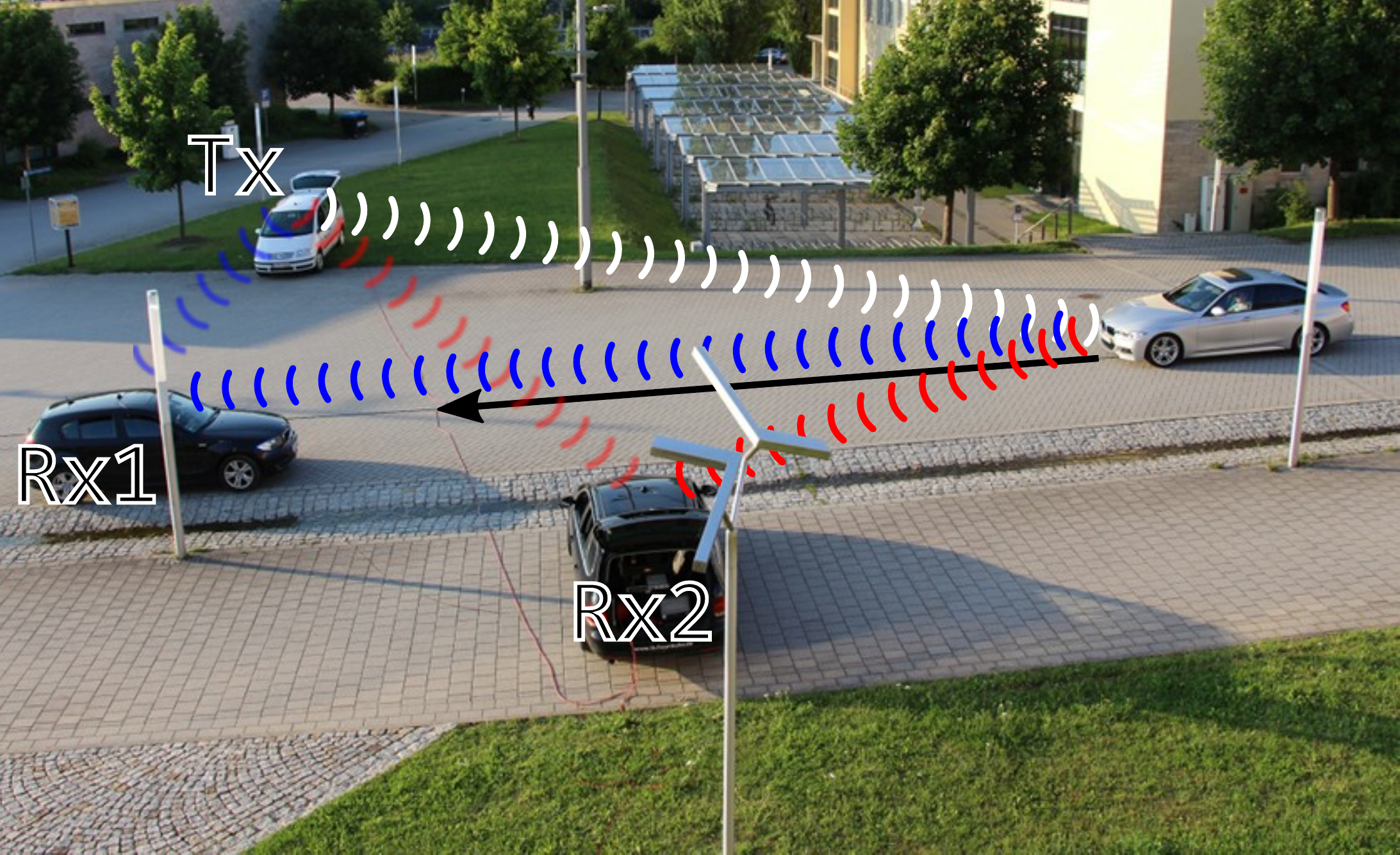}%
        \label{fig_meas_setup}
    }\\
    \centering
    \vspace{0.5em}
    \includegraphics[width=.95\columnwidth]{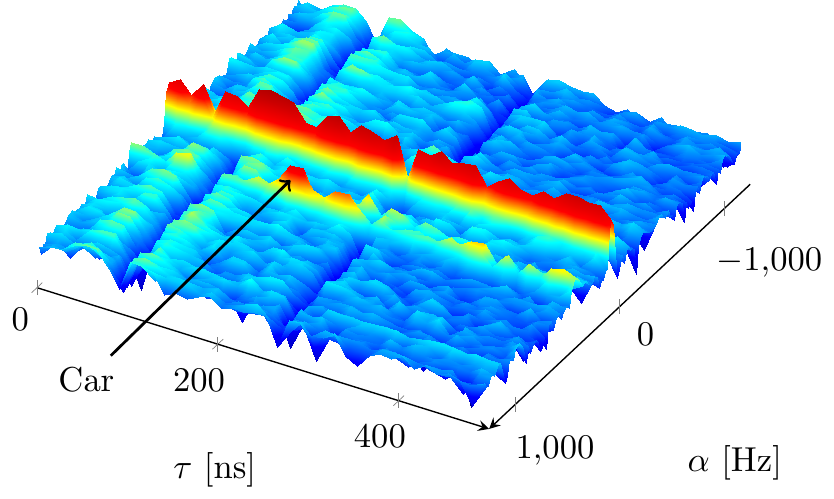}\\
    \vspace{-1em}
    \subfloat[]{
        \centering
        \includegraphics[width=.8\columnwidth]{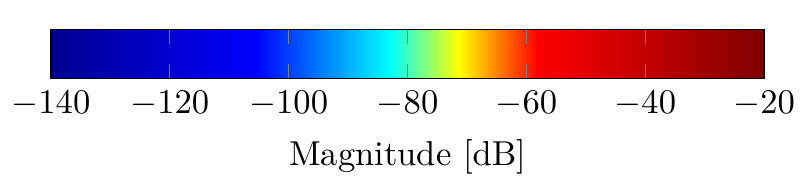}%
        \label{fig_meas_Dft3}
    }
    \caption{
        Measurement setup and results.
        a)
        Bi-static radar scenario.
        Transmitting and receiving cars are static.
        The radar target on the right drives along the black arrow.
        b)
        Doppler shift ($\alpha$) vs. fast time ($\tau$) response at Rx 2.
        The scattering function clearly indicates the moving target.
        The static clutter collapses at Doppler shift zero.
        Both visible rifts stem from removal of strong static paths via high-resolution post-processing.
    }
    \label{fig_meas}
\end{figure}

In the following, we present first results from an initial measurement campaign of a bi-static radar scenario with one illuminator (Tx) and two sensors (Rx) as shown in Fig. \ref{fig_meas_setup}.
In this simple example, Tx and both Rx were stationary and only the target car was moving.

The goal of the experiment was to demonstrate the feasibility of the signal processing approach described in Fig. 3 and to answer the question: “Given typical radio parameters and a typical bi-static road traffic scenario, can we see the moving object?”

The measurement was carried out with spatially distributed software-defined radio modules as transceivers and an \SI{80}{MHz} \gls{OFDM} signal.
The slow-time sequence of the power delay profiles (not shown here) does not reveal the target as it is masked by the strong \gls{LoS} signal and static clutter.
However, it is clearly visible in the magnitude-squared delay-Doppler responses (scattering functions) in Fig.~\ref{fig_meas_Dft3}, which are calculated by a \SI{10}{ms} \gls{FFT} (\gls{D-FFT}) along slow time at sensor Rx2.
The scattering function at Rx1 would show the target response at a different time delay and a different Doppler shift according to the respective bi-static geometry.

This example clearly shows that clutter removal would need well-defined signal processing measures, which exploit the dedicated Doppler shift of the target to distinguish it from the static background.
High-resolution parameter estimation can further enhance target location and clutter removal.

\begin{table}[h]
    \begin{center}
        \caption{Overview comparison of \gls{CPCL} to passive radar (PCL) and conventional dedicated radar.}
        \label{tab:table1}
        \begin{tabular}{lccc} % <-- Changed to S here.
            \toprule
            Feature & \gls{CPCL} & \gls{PCL} & Dedicated radar\\
            \midrule
            Range resolution & + & 0 & ++\\
            Diversity gain & ++ & + & 0 \\
            Low signal processing effort & + & + & ++ \\
            Resource efficiency & ++ & + & 0 \\
            Level of cooperation & ++ & 0 & 0 \\
            Coverage and ubiquitousness & ++ & + & 0 \\
            Standalone operation & 0 & 0 & ++ \\
            Integrated service level & ++ & 0 & 0 \\
            Adaptability & ++ & 0 & + \\ \bottomrule
        \end{tabular}
    \end{center}
\end{table}

\section*{Conclusion and Outlook}
We have introduced and explained the new concept of \gls{CPCL}.
Within the \gls{5G} perspective, \gls{CPCL} promises to turn the mobile radio network into an ubiquitous radar network, which may be adaptive, reconfigurable, and even cognitive.
The scalability and flexibility of \gls{5G} will allow tailoring of \gls{CPCL} to a variety of application classes.
The real-time computing facilities of the \gls{MEC} will support radar data fusion on sensor and scenario levels and eventually enable the \glspl{MNO} to offer \gls{CPCL} as an integrated service for public user groups.
The unprecedented service potential of \gls{CPCL} comes from the fact that it exploits the newest radio communication principles developed throughout an unrestrained progress in mobile radio over the last decades.
As \gls{CPCL} makes secondary use of communication signals and network resources for radar, it can be seen as a resource-saving green technology.
This goes along with the access to all radio frequencies assigned for mobile services which not only opens a huge potential for radar frequency diversity but may even solve the competition issue in frequency assignment between radar and communication community.
Whereas dedicated automotive radar at millimeter-wave frequencies currently has the highest range resolution potential, \gls{CPCL} will gain with the 5G millimeter-wave bands. Moreover, carrier aggregation at frequencies below 6 GHz effectively offers GHz resolution capability together with the range advantage of the lower frequencies.

\gls{CPCL} is a comprehensive integration of radar functionalities into the framework of mobile radio systems.
This makes the difference to the alternative \emph{RadCom} idea \cite{5776640}, which assumes a communication link extension to radar.
Therefore, we propose the term \emph{ComRad} as another appropriate acronym for \gls{CPCL}.

Compared to simple passive radar, \gls{CPCL} has many advantages, which we have discussed in this paper (see also Table 1 for an overview).
Among those are a full synchronous and orthogonal operation (also for multiple sensors) which reduces estimation variance.
This cannot be achieved with passive radar as \gls{PCL} lacks any cooperation between the illuminating and sensing radio nodes. Therefore, \gls{PCL} does not allow adaptive radio resource allocation and for data fusion it needs a separate communication network.
On the other hand, \gls{CPCL} allows a high level of cooperation since both the illuminator and the sensor are booked in the same network.

The advantage of \gls{CPCL} to conventional automotive radar is that a \gls{CPCL} network may automatically have access to all the communication and data fusion capabilities at hand.
This allows building synchronous \gls{SIMO}, \gls{MISO}, or \gls{MIMO} radar networks, which would be necessary to reach full road traffic situation awareness on scenario levels including multi-lane crossings and spread out traffic hotspots.
The bi-static view, the dense network of sensors, frequency diversity and adaptive/cognitive management of the available radio resources offer unprecedented performance features.
Furthermore, synchronous signaling and medium access control schemes, which are an inherent part in a \gls{CPCL} network, will automatically solve many interference and collision problems that conventional dense radar networks will be faced with in the future.
This brings us to the obvious question: Can radar borrow ideas for medium access control and radio resources scheduling from mobile radio? \gls{CPCL} is the comprehensive and positive answer to this question.

\gls{CPCL} can actually be seen as a framework that supports and extends conventional automotive radar. \gls{CPCL} may even host stand-alone radar as an auxiliary class of sensors in a generic communication centered ComRad radar network.

Although the dictum of this paper was driven by requirements from the automotive sector, it becomes obvious that there may be other vertical markets related to mobility, security, and industrial areas, where an integrated communications and radar service could be a great benefit.
We eventually deem that \gls{CPCL} could be a service that may be offered by the \glspl{MNO} to public user groups and public safety agencies, for example, for road traffic monitoring, logistics, mobility, and several security applications, as it is likely that \gls{5G} networks will play a more important role for safety and mission-critical communication than earlier mobile radio generations \cite{7890058}.

% or


%\appendices

%\section{}

%\section*{Acknowledgment}
%This work was supported by the Freistaat Thüringen and the European Social Fund.
\ifCLASSOPTIONcaptionsoff
  \newpage
\fi



\bibliographystyle{IEEEtran}
\bibliography{comsoc_cpcl}
%\begin{thebibliography}{1}
%\end{thebibliography}


\vskip -1\baselineskip plus -1fil
\begin{IEEEbiographynophoto}{Reiner S. Thomä}
[M'92, SM'00, F'07] (reiner.thomae@tu-ilmenau.de) received his degrees in electrical engineering and information technology from TU Ilmenau, Germany, where he has been a full professor since 1992.
In 2014 he received the Vodafone Innovation Award.
\end{IEEEbiographynophoto}
\vskip -2.2\baselineskip plus -1fil
\begin{IEEEbiographynophoto}{Carsten Andrich} 
(carsten.andrich@iis.fraunhofer.de) is currently a researcher with the Fraunhofer Institute for Integrated Circuits IIS and a Ph.D. candidate specializing in SDR applications.
\end{IEEEbiographynophoto}
\vskip -2.2\baselineskip plus -1fil
\begin{IEEEbiographynophoto}{Giovanni Del Galdo}
[M'12] (giovanni.delgaldo@iis.fraunhofer.de) studied telecommunications engineering at Politecnico di Milano.
Since 2012 he is full professor at TU Ilmenau, now leading a joint research group of TU Ilmenau and Fraunhofer IIS counting 45 researchers.
\end{IEEEbiographynophoto}
\vskip -2.2\baselineskip plus -1fil
\begin{IEEEbiographynophoto}{Michael Döbereiner}
(michael.doebereiner@iis.fraunhofer.de) is currently a researcher with the Fraunhofer Institute for Integrated Circuits IIS and a Dr.\hbox{-}Ing. candidate with research focus on high resolution parameter estimation in dynamic scenarios.
%received the B.Sc. and M.Sc. degrees in electrical engineering and information technology from the Technische Universität Ilmenau in 2017 and 2018.
%He is currently working at the Fraunhofer Institute for Integrated Circuits IIS.
%His research interests include passive radar technologies and the high resolution parameter estimation of radar signals.
\end{IEEEbiographynophoto}
\vskip -2.2\baselineskip plus -1fil
\begin{IEEEbiographynophoto}{Matthias A. Hein}
[M'06, SM'06] (matthias.hein@tu-ilmenau.de) joined the TU Ilmenau in 2002 as Head of the RF~{\&} Microwave Research Group. His current research focuses on automotive wireless sensor, communication, and navigation systems and virtual test drives.
\end{IEEEbiographynophoto}
\vskip -2.2\baselineskip plus -1fil
\begin{IEEEbiographynophoto}{Martin Käske}
(martin.kaeske@tu-ilmenau.de) received the Dipl.-Ing. degree in electrical engineering (information technology) from Technische Universit\"at Ilmenau, Germany. His research work is focused on high resolution parameter estimation.
%received the Dipl.-Ing. (M.S.E.E.) degree in electrical engineering (information technology) from Technische Universität Ilmenau, Germany, in 2008.
%He is currently working at the Electronic Measurement Research Lab, TU Ilmenau.
%He is focusing on diffuse scattering—its modeling and estimation—as well as high resolution parameter estimation.
\end{IEEEbiographynophoto}
\vskip -2.2\baselineskip plus -1fil
\begin{IEEEbiographynophoto}{Steffen Schieler}
(steffen.schieler@tu-ilmenau.de) is currently a researcher with the Electronic Measurement and Signal Processing Group at the Technische Universit\"at Ilmenau, Germany, and Ph.D. student with a focus on signal processing.
% obtained his M.Sc. in Electrical Engineering and ICT from Technische Universit\"at Ilmenau, Germany, in 2018.
% He is currently working at the Electronic Measurements and Signal Processing Lab at Technische Universit\"at Ilmenau, Germany.
% His research interest includes passive localization techniques and next-generation wireless communication systems.
\end{IEEEbiographynophoto}
\vskip -2.2\baselineskip plus -1fil
\begin{IEEEbiographynophoto}{Christian Schneider}
(christian.schneider@tu-ilmenau.de) received his Diploma in electrical engineering from the Technische Universität Ilmenau in 2001. Currently he is a senior researcher and program manager at the Technische Universität Ilmenau.
% received his Diploma degree in electrical engineering from the Technische Universität Ilmenau, Germany in 2001.
% His research interests include multi-dimensional channel sounding, characterization and modelling for single and multi-link cases in cellular and vehicular networks at micro and millimeter wave bands.
% He is also active in adaptive space-time signal processing and passive coherent localisation.
% He received a best paper award at the European Wireless conference in 2013 and European Conference of Antennas and Propagation in 2017.
\end{IEEEbiographynophoto}
\vskip -2.2\baselineskip plus -1fil
\begin{IEEEbiographynophoto}{Andreas Schwind}
(andreas.schwind@tu-ilmenau.de) is currently a researcher with the RF and Microwave Research Group at the Technische Universit\"at Ilmenau, Germany, and a Dr.\hbox{-}Ing. candidate specializing in bi-static radar cross-section measurements.
% received his M.Sc. degree in electrical engineering from the Technische Universit\"at Ilmenau, Germany, in 2017.
% He is currently working in the RF and Microwave Research Group at the same university.
\end{IEEEbiographynophoto}
\vskip -2.2\baselineskip plus -1fil
\begin{IEEEbiographynophoto}{Philip Wendland}
(philip.wendland@tu-ilmenau.de) is currently a Ph.D. candidate with the Group for Telematics and Computer Networks at the Technische Universität Ilmenau.
His research focus is on medium access control for vehicular communication.
\end{IEEEbiographynophoto}
\vskip -2.2\baselineskip plus -1fil
\begin{IEEEbiographynophoto}{Günter Schäfer}
[M'95] (guenter.schaefer@tu-ilmenau.de) received his Ph.D. in computer science from the University of Karlsruhe in 1998.
After researching and teaching at ENST in Paris and TU Berlin, he was appointed full professor at the TU Ilmenau in 2005.
\end{IEEEbiographynophoto}

\newpage


%\vfill


\end{document}